\documentstyle[aps,prl,psfig]{revtex}
\input psfig
\begin{document}
\draft
\twocolumn[\hsize\textwidth\columnwidth\hsize\csname @twocolumnfalse\endcsname
\title{Multiple light scattering in nematic liquid crystals}
\author{Holger Stark$^{*}$ and Tom C. Lubensky}
\address{Department of Physics and Astronomy, University of Pennsylvania,
Philadelphia, PA 19104, USA}
\maketitle
\begin{abstract}
We present a rigorous treatment of the diffusion approximation for multiple 
light scattering in anisotropic random media, and apply it to director
fluctuations in a nematic liquid crystal. For a typical nematic material,
5CB, we give numerical values of the diffusion constants $D_{\|}$ and 
$D_{\perp}$.
We also calculate the temporal autocorrelation function measured in Diffusing
Wave Spectroscopy.
\end{abstract}
\vskip2pc]
\narrowtext

Light transport in random or turbid media has long been treated by 
radiative transfer theories, the first of which was formulated  as
early as 1905 by Schuster \cite{Schuster05}. For distances large compared to
the transport mean-free-path $l^{\ast}$, beyond which the direction of 
light propagation is randomized, these theories can be reduced 
\cite{Ishimaru78} to a diffusion equation for the light energy density
with diffusion constant $D=\bar{c}l^{\ast}/3$ where $\bar{c}$ is the speed 
of light in the medium. In 1984 Kuga and Ishimaru
\cite{Kuga84} discovered coherent backscattering of light in colloidal
suspension, predicted in earlier
papers \cite{Golubentsev84}, and physicists realized the connection
of wave propagation in disordered media to weak localization 
\cite{Wolf85}, a precursor of Anderson localization 
\cite{Anderson58}. Since then, our theoretical understanding of light 
transport in random media has advanced considerably. Detailed studies of 
multiple scattering of scalar waves \cite{Sheng90} was followed by the
generalization to include the polarization of light
\cite{Stephen86}, broken time reversal symmetry and optical
activity \cite{Golubentsev84,MacKintosh88}, and long-range correlations
in random scatterers \cite{John83,MacKintosh89}.
Multiple scattering emerged as a powerful probe of
dynamical properties of turbid media with the development of Diffusing
Wave Spectroscopy (DWS) \cite{Maret87,Pine88} as an experimental technique 
capable of measuring dynamic correlations at time scales, much shorter
\cite{Kao93}, than can be probed with single scattering.

\begin{figure}
\centerline{\psfig{figure=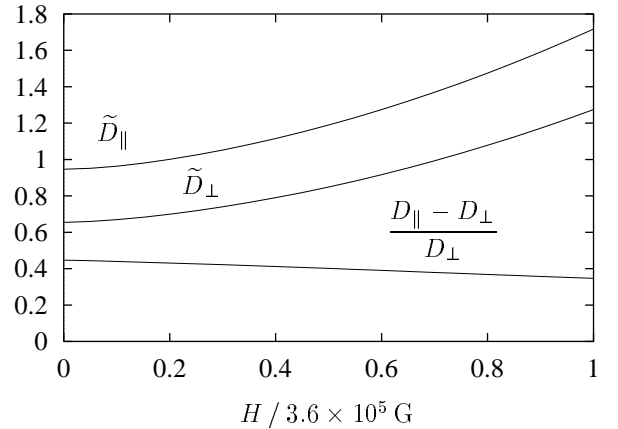}}
    \caption[]{The field dependence of the normalized diffusion constants 
 $\widetilde{D}_{\|}$ and $\widetilde{D}_{\perp}$ and the anisotropy 
 $(D_{\|}-D_{\perp})/D_{\perp}$ for parameters of a typical nematic
 liquid crystal 5CB: $K_1/K_3=0.79$, $K_2/K_3=0.43$ and 
 $\Delta \varepsilon / \varepsilon_{\perp} = 0.228$.}
    \label{fig:1}
\end{figure}

Nematic liquid crystals are strong light scatterers, exhibiting turbidity
and coherent backscatter \cite{Vithana93}. They differ, however, in 
significant ways from colloidal suspensions, the most widely studied 
multiple-scattering media. First, nematic liquid crystals are anisotropic
with barlike molecules aligned on average along a unit vector 
$\bbox{n}(\bbox{r},t)$ called the director. They are birefringent with 
different velocities of light for ordinary and extraordinary rays. As a 
result the photon energy density, like particle density in an electron
system \cite{Woelfle84}, obeys an anisotropic diffusion equation with 
diffusion coefficients $D_{\|}$ and $D_{\perp}$ for directions parallel
and perpendicular to the equilibrium director $\bbox{n}_0$. Second, the
dominant scattering of visible light is from {\em long-range\/} thermal 
fluctuations of the director rather than from particles with diameters
comparable to the wavelength of light. This leads to a divergent scattering
mean-free-path when the external magnetic field $H$ is zero
\cite{Val'kov86,Langevin75}. The diffusion constants $D_{\|}$ and $D_{\perp}$,
which in isotropic systems are proportional to the transport mean-free-path,
are, however, finite when $H \rightarrow 0$ as shown in Fig. \ref{fig:1}.
In this paper, we develop a systematic
treatment of the diffusion approximation for multiple light scattering
in anisotropic random media (for recent approaches see \cite{Tiggelen95}),
which allows us to calculate $D_{\|}$ and $D_{\perp}$ from known properties
of nematics and to obtain the time-dependent response measured in DWS
experiments\cite{Stephen88}.  
Fig.\ \ref{fig:1} shows our calculated values of $D_{\|}$
and $D_{\perp}$ as a function of external magnetic field $H$ for the 
compound 5CB. The anisotropy ratio $D_{\|}/D_{\perp}=1.45$ is in good
agreement with measurements reported in a companion paper \cite{Yodh95}.

We start with the wave equation for the electric light field
$\bbox{E}(\bbox{r},t)$:
\begin{equation}
\label{1}
\Bigl\{ \bbox{\nabla} \times \bbox{\nabla} \times +\frac{1}{c^2}\,
\frac{\partial^2}{\partial t^2} \, [\bbox{\varepsilon}_0 +
\delta \bbox{\varepsilon}(\bbox{r},t)]\Bigr\} \, 
\bbox{E}(\bbox{r},t) = \bbox{0} 
\enspace .
\end{equation}
The homogeneous part of the dielectric tensor is $\bbox{\varepsilon}_0$, and
the randomly fluctuating part $\delta \bbox{\varepsilon}(\bbox{r},t)$ is 
a Gaussian random variable described by the correlation function
\begin{equation}
\label{2}
\bbox{B}^{\omega}(\bbox{R},t):= \frac{\omega^4}{c^4}\,
\langle \, \delta \bbox{\varepsilon}(\bbox{R},t)) \otimes 
\delta \bbox{\varepsilon}(\bbox{0},0) \, \rangle^{(N)} \enspace ,
\end{equation}
where $\omega$ is the frequency of light. The superscript $(N)$ means that we 
interchange the second and third index in the tensor product
$\delta \bbox{\varepsilon} \otimes \delta \bbox{\varepsilon}$ to
define $\bbox{B}^{\omega}$: $[\bbox{B}^{\omega}]_{ijkl} \propto \langle \,
\delta \varepsilon_{ik} \, \delta \varepsilon_{jl} \, \rangle$. We call 
$\bbox{B}^{\omega}(\bbox{R},t)$ the structure factor of the
system. It is measured in single light scattering experiments
\cite{Berne76} and contains informations about the elastic and dynamic
properties of a system.  
The local uniaxial dielectric tensor can be expressed as
\begin{equation}
\label{3}
\bbox{\varepsilon}(\bbox{r},t)=\varepsilon_{\perp}\bbox{1} + \Delta 
\varepsilon [\bbox{n}(\bbox{r},t) \otimes \bbox{n}(\bbox{r},t)] 
\end{equation}
Here $\varepsilon_{\perp}$ and
$\varepsilon_{\|}$ are the dielectric constants for electric fields,
respectively, perpendicular and parallel to the director, and 
$\Delta \varepsilon=\varepsilon_{\|}-\varepsilon_{\perp}$. We assume
that the inhomogeneity of the director field only comes from thermal 
fluctuations of the director around its equilibrium value $\bbox{n}_0$:
$
\bbox{n}(\bbox{r},t) = \bbox{n}_0 + \delta\bbox{n}(\bbox{r},t)
$,
where $\delta \bbox{n}$ has to be perpendicular to $\bbox{n}_0$ for
small fluctuations. 
The dominant contribution to $\bbox{B}^{\omega}(\bbox{R},t)$ is proportional
to the director correlation function 
$\langle \,\delta\bbox{n}(\bbox{R},t) \otimes 
\delta\bbox{n}(\bbox{0},0) \,\rangle$ which we express in momentum space 
\cite{Gennes75}:
\begin{eqnarray}
\label{5}
\lefteqn{ \langle \,\delta\bbox{n}(\bbox{q},t) \otimes
\delta\bbox{n}^{\ast}(\bbox{q},0) \,\rangle = }\qquad \nonumber\\ & & 
\sum_{\alpha=1}^{2} \frac{k_{\text{B}}T}{K_{\alpha}(\bbox{q})} \, 
\exp\left[-\frac{K_{\alpha}(\bbox{q})}{\eta_{\alpha}(\bbox{q})} \,t \right]
\, \hat{\bbox{u}}_{\alpha}(\bbox{q}) \otimes \hat{\bbox{u}}_{\alpha}(\bbox{q}) 
\enspace.
\end{eqnarray}
Here $ K_{\alpha}(\bbox{q})= K_{\alpha} q_{\perp}^2 + K_3 q_{\|}^2 + \Delta
\chi H^2 $, where $K_1$, $K_2$, and $K_3$ are the Frank elastic constants,
$H$ is the external field parallel to $\bbox{n}_0$, and $\Delta \chi$ is the
anisotropy of the magnetic susceptibility. The quantity 
$\eta_{\alpha}(\bbox{q})$ is a combination
of viscosities which appear in the hydrodynamic equations of the 
director field, the Leslie-Erickson equations \cite{Gennes75}. The unit
vectors $\hat{\bbox{u}}_{\alpha}(\bbox{q})$ specify the direction of
$\delta\bbox{n}(\bbox{q},t)$ in mode $\alpha=1$ and 2.

In an anisotropic medium with a homogeneous dielectric tensor
$\bbox{\varepsilon}_0$ the electromagnetic field travelling along the
unit vector $\hat{\bbox{k}}$ has two modes with indices of refraction
$n_{\alpha}(\hat{\bbox{k}})$ and electric polarization 
$\bbox{e}_{\alpha}(\hat{\bbox{k}})$.
The polarization $\bbox{d}^{\alpha}(\hat{\bbox{k}})$ of the displacement 
field, $\bbox{d}^{\alpha}(\hat{\bbox{k}}) = \bbox{\varepsilon}_0 
\bbox{e}_{\alpha}(\hat{\bbox{k}})$, obeys $\bbox{d}^{\alpha}(\hat{\bbox{k}})
\cdot \hat{\bbox{k}} =0$. After an appropriate
normalization, the vectors fulfill the biorthogonality relation 
$\bbox{d}^{\alpha}(\hat{\bbox{k}}) \cdot \bbox{e}_{\beta}(\hat{\bbox{k}})
 = \delta^{\alpha}_{\beta}$ \cite{Lax71}. We can now write down the
momentum space representation of the averaged retarded and advanced
Green's function of Eq.~(\ref{1}) in the weak-scattering approximation:
\begin{eqnarray}
\langle\, \bbox{G}^{R/A} \,\rangle(\bbox{k},\omega) & \approx &
\sum_{\alpha=1}^{2} 
\Bigl\{ \Bigl( \omega/c \pm i/[2n_{\alpha}(\hat{\bbox{k}}) 
l_{\alpha}(\hat{\bbox{k}},\omega)] \Bigr)^2  \nonumber\\
\label{7}
 & &  - \, k^2/n_{\alpha}^2(\hat{\bbox{k}}) \Bigr\}^{-1} \,
\bbox{e}_{\alpha}(\hat{\bbox{k}}) \otimes \bbox{e}_{\alpha}(\hat{\bbox{k}})
\end{eqnarray}
with
\begin{eqnarray}
\label{8}
l_{\alpha}(\hat{\bbox{k}},\omega) =  \biggl[\frac{\pi}{2} \,
n_{\alpha}(\hat{\bbox{k}}) \sum_{\beta=1}^{2} \int_{\hat{\bbox{q}}^{\beta}} 
[\bbox{B}^{\omega}_{\bbox{k}^{\alpha}\bbox{q}^{\beta}}(t=0)]_{\alpha\beta}
 \biggr]^{-1}
\end{eqnarray}
being the scattering mean-free-path of a light mode 
$\{\bbox{k}^{\alpha}\,|\, \bbox{e}_{\alpha}(\hat{\bbox{k}}) \}$ with wave 
vector $\bbox{k}^{\alpha} = \frac{\omega}{c}n_{\alpha} \hat{\bbox{k}}$ and
polarization $\bbox{e}_{\alpha}(\hat{\bbox{k}})$. The structure factor 
$[\bbox{B}^{\omega}_{\bbox{k}^{\alpha}\bbox{q}^{\beta}}
(t=0)]_{\alpha\beta}$ describes the scattering of 
$\{ \bbox{k}^{\alpha}\, | \, \bbox{e}_{\alpha}(\hat{\bbox{k}}) \}$ into 
$\{ \bbox{q}^{\beta} \, | \, \bbox{e}_{\beta}(\hat{\bbox{k}}) \}$.
The symbol $\int_{\hat{\bbox{q}}^{\beta}}$ always stands for an angular
integration:
\begin{equation}
\int_{\hat{\bbox{q}}^{\beta}} \ldots \enspace = 
\int \frac{d\Omega_{\bbox{q}}}{(2\pi)^3}
n_{\beta}^3(\hat{\bbox{q}}) \,\, \ldots 
\enspace.
\end{equation}
To the order of our calculations 
$\langle\, \bbox{G}^{R/A} \,\rangle(\bbox{k},\omega)$ is diagonal in the
polarization $\bbox{e}_{\alpha}(\hat{\bbox{k}})$. In what follows, a Greek
index will refer to the ``basis vectors'' 
$\bbox{e}_{\alpha} \otimes \bbox{e}_{\alpha}$ or 
$\bbox{d}^{\alpha} \otimes \bbox{d}^{\alpha}$.
The scattering mean-free-path $l_{\alpha}(\hat{\bbox{k}},\omega)$ in the
nematic phase has been calculated \cite{Val'kov86,Langevin75}. 
For small $H$ for the extraordinary ray it has the
form $l_{\alpha}^{-1} \propto \bigl(\frac{\omega}{c}\bigr)^2 
\frac{k_{\text{B}}T}{K}\text{ln}\bigl(\frac{\Delta \chi H^2}{K(\omega/c)^2}
\bigr)$, where $K$ is an appropriately averaged elastic constant.

Let us look at the spatial and temporal autocorrelation function for
the electric light field: $\langle \, \bbox{E}(\bbox{R}+\frac{\bbox{r}}{2},
 T+\frac{t}{2}) \otimes \bbox{E}^{\ast}(\bbox{R}-\frac{\bbox{r}}{2}, 
T-\frac{t}{2}) \, \rangle$, where we have already introduced center of ``mass''
($\bbox{R},T$) and relative ($\bbox{r},t$) coordinates. From this
quantity, others follow as special cases: the energy density of
light at time $T$ is $W_1(\bbox{R},T)=\langle \, \bbox{E}(\bbox{R}, T) \cdot
\bbox{\varepsilon}_0 
\bbox{E}^{\ast}(\bbox{R},T) \, \rangle$, where the $T$ dependence is, e.g.,
due to time dependent-sources; the temporal correlation function of
a steady-state light field is $W_2(\bbox{R},t) =\langle \, 
\bbox{E}(\bbox{R},\frac{t}{2}) \cdot \bbox{\varepsilon}_0
\bbox{E}^{\ast}(\bbox{R},-\frac{t}{2}) \, \rangle$, which reflects the
dynamics of the scattering media measured in DWS experiments.
The Fourier transform with respect
to $\bbox{r}$ gives the energy density with wave vector $\bbox{k}$ 
\cite{Ishimaru78}. 
To calculate the autocorrelation function
for special light sources and/or given boundary conditions, we need the
``two-particle'' Green's function
$ \bbox{\Phi} = \langle \, \bbox{G}^{R} \otimes
\bbox{G}^{A} \, \rangle^{(N)}$.
Our goal is to derive the diffusion pole of $\bbox{\Phi}$ in momentum and 
frequency space. With all arguments, the Green's function
is $\bbox{\Phi}^{\omega}_{\bbox{k}\bbox{k}'}(\bbox{K},\Omega,t)$. 
$\bbox{K}$, $\Omega$ correspond to the center of ``mass''
coordinates $\bbox{R}$, $T$ and $\bbox{k}$, $\bbox{k}'$ to the 
relative coordinates $\bbox{r}$,$\bbox{r}'$. The superscript $\omega$
is the light frequency, and the $t$ dependence explicitly comes from the
structure factor $\bbox{B}^{\omega}_{\bbox{k}\bbox{k}'}(t)$.
In the weak-scattering approximation, 
$\bbox{\Phi}^{\omega}_{\bbox{k}\bbox{k}'}(\bbox{K},\Omega,t)$ can be 
represented as a sum of ladder diagrams, which is equivalent to the 
Bethe-Salpeter equation:
\begin{eqnarray}
&\displaystyle
 \int \frac{d^3k_1}{(2\pi)^3} [\bbox{1}_{\bbox{k}\bbox{k}_1}^{(4)} - 
\bbox{f}^{\omega}_{\bbox{k}}(\bbox{K},\Omega)\,
\bbox{B}^{\omega}_{\bbox{k}\bbox{k}_1}(t)] 
\bbox{\Phi}^{\omega}_{\bbox{k}_1\bbox{k}'}(\bbox{K},\Omega,t) & \nonumber\\
\label{10}
& = \bbox{f}^{\omega}_{\bbox{k}}(\bbox{K},\Omega) 
\bbox{1}_{\bbox{k}\bbox{k}'}^{(4)} \enspace,
\end{eqnarray}
where
\begin{equation}
\label{11}
\bbox{f}^{\omega}_{\bbox{k}}(\bbox{K},\Omega) =  \bigl[\langle
\bbox{G}^{R}\rangle (\bbox{k}_{+} ,\omega_{+} )  \otimes\, 
\langle \bbox{G}^{A} \rangle (\bbox{k}_{-},\omega_{-}) \bigr]^{(N)} 
\end{equation}
with $\bbox{k}_{\pm}=\bbox{k}\pm \bbox{K}/2$ and 
$\omega_{\pm}=\omega \pm \Omega/2$
and,
$
[\bbox{1}_{\bbox{k}\bbox{k}'}^{(4)}]_{ijkl} := (2\pi)^3 \,
\delta(\bbox{k}-\bbox{k}')\,(\delta_{ik}\delta_{jl}+\delta_{il}
\delta_{jk}) / 2
$.
The multiple integrals and the sum can be done 
analytically for $\delta$-function correlations but not for the anisotropic,
long-range correlations of our problem.
Vollhardt and W\"olfle \cite{Vollhardt80} derived the diffusion pole for
isotropic electron transport directly from Eq.\ (\ref{10}), and 
MacKintosh and John \cite{MacKintosh89} applied their method to light.
In the anisotropic case one has to be more careful \cite{Woelfle84}.
If $\bbox{\Psi}_{\bbox{k}}^{(n)}(\bbox{K},\Omega,t)$
and $\lambda^{(n)} (\bbox{K},\Omega,t)$ are, respectively, the $n$th 
eigenvector and eigenvalue of the integral operator of Eq.\ (\ref{10}),
\begin{equation}
\label{13}
 \int \frac{d^3k_1}{(2\pi)^3} [\bbox{1}_{\bbox{k}\bbox{k}_1}^{(4)} - 
\bbox{f}^{\omega}_{\bbox{k}}(\bbox{K},\Omega)\,
\bbox{B}^{\omega}_{\bbox{k}\bbox{k}_1}(t)]\bbox{\Psi}_{\bbox{k}_1}^{(n)} 
= \lambda^{(n)} \bbox{\Psi}_{\bbox{k}}^{(n)} \enspace,
\end{equation}
and $\overline{\bbox{\Psi}}_{\bbox{k}}^{(n)}(\bbox{K},\Omega,t)$ the 
eigenvectors of the Hermitian adjoint operator,
it is straightforward to show that
\begin{equation}
\label{14}
\bbox{\Phi}^{\omega}_{\bbox{k}\bbox{k}'}(\bbox{K},\Omega,t) = \sum_{n}
\frac{\bbox{\Psi}_{\bbox{k}}^{(n)} \otimes 
\overline{\bbox{\Psi}}_{\bbox{k}'}^{(n)}}{\lambda^{(n)} 
} \, \bbox{f}^{\omega}_{\bbox{k}'}(\bbox{K},\Omega)
\end{equation}
solves the Bethe-Salpeter equation \cite{MacKintosh89}.
In the case of $\bbox{K}=\bbox{0}$ and $\Omega=t=0$, it can be shown that
the quantity $\Delta\bbox{G}_{\bbox{k}}^{\omega}(\bbox{0},0)$, where
\begin{equation}
\label{15}
\Delta\bbox{G}_{\bbox{k}}^{\omega}(\bbox{K},\Omega) =  
\langle \bbox{G}^{R}\rangle \bigr(\bbox{k}_{+}, \omega_{+} \bigl)
- \langle \bbox{G}^{A}\rangle \bigl(\bbox{k}_{-}, \omega_{-} \bigr) \enspace,
\end{equation}
is an eigenvector with eigenvalue $\lambda^{(0)}(\bbox{0},0,0) = 0$. This
is a very general result, based on the Ward identities valid beyond
the weak-scattering approximation \cite{Vollhardt80}. We have identified
the diffusion pole, as we shall explicitly see soon. All other eigenvalues 
are positive, and, in real space, they give exponentially decaying 
contributions 
to $\bbox{\Phi}^{\omega}_{\bbox{k}\bbox{k}'}(\bbox{K},\Omega,t)$ of 
Eq.\ (\ref{14}) , which are not important at long length scales 
\cite{MacKintosh89}. To establish the diffusion approximation we have to
apply perturbation theory to calculate $\lambda^{(0)}(\bbox{K},\omega,t) $
for small $\bbox{K}$, $\Omega$, and $t$. Therefore, we expand the 
eigenvectors into a set of basis functions and turn the eigenvalue equation
(\ref{13}) into a matrix equation. For the component $\alpha$ of 
$\bbox{\Psi}_{\bbox{k}}$
we use the ansatz
\begin{equation}
\label{16}
\Psi^{\alpha}_{\bbox{k}} \propto
[\Delta \bbox{G}_{\bbox{k}}^{\omega}(\bbox{0},0)]^{\alpha} \,
 \bigl[\widetilde{\Psi}_0 \pi + 
\sum_{i} 
\widetilde{\Psi}_{i}^{\alpha} \, \varphi_{i}^{\alpha}(\hat{\bbox{k}})
\bigr]
\enspace,
\end{equation}
where
\begin{equation}
\label{17}
[\Delta\bbox{G}_{\bbox{k}}^{\omega}(\bbox{0},0)]^{\alpha} \approx
 -i\pi\frac{c}{\omega} n_{\alpha}(\hat{\bbox{k}}) \,
 \delta\Bigl(\frac{\omega}{c} n_{\alpha}(\hat{\bbox{k}})-k\Bigr) \enspace.
\end{equation}
The first factor on the right-hand side of Eq.\ (\ref{16}) is due to the 
momentum shell 
approximation, Eq.\ (\ref{17}); it is strongly peaked 
around the wave numbers of the light modes. The amplitude $\widetilde{\Psi}_0$
then represents the zeroth eigenvector. The second term corresponds to the
space of all other eigenvectors where the 
$\varphi_{i}^{\alpha}(\hat{\bbox{k}})$ are
basis functions on the unit sphere, e.g. spherical harmonics, which can in
general depend on polarization $\alpha$.
We also use the relation
\begin{eqnarray}
\lefteqn{[\Delta \bbox{G}_{\bbox{k}}^{\omega}(\bbox{K},\Omega)]^{\alpha} 
\approx [\bbox{f}^{\omega}_{\bbox{k}}(\bbox{0},0)]^{\alpha \alpha}} 
 \qquad \quad    \nonumber\\
\label{18}
 & & \times \left[ \Delta \bbox{\Sigma}_{\bbox{k}}^{\omega}(\bbox{0},0) -
\frac{\partial \bbox{G}_0^{-1}}{\partial \bbox{k}} \bbox{K} - 
\frac{\partial \bbox{G}_0^{-1}}{\partial \omega} \Omega \right]_{\alpha}
\enspace,
\end{eqnarray}
which gives $[\Delta \bbox{G}_{\bbox{k}}^{\omega}(\bbox{K},\Omega)]^{\alpha}$
correctly to first order in $\bbox{K}$, $\Omega$ and $\bbox{\Sigma}$ 
(see \cite{Stark95c}). ($\bbox{\Sigma}$ is the mass operator and 
$\Delta \bbox{\Sigma}_{\bbox{k}}^{\omega}$ is defined the same way as
$\Delta \bbox{G}_{\bbox{k}}^{\omega}$. $\bbox{G}_0$ stands for the Green's
function of the homogeneous medium.) 
The coupling between the zeroth and the other eigenvectors then produces the 
diffusion tensor, and, finally, the Green's function takes the form:
\begin{equation}
\bbox{\Phi}^{\omega}_{\bbox{k}\bbox{k}'}(\bbox{K},\Omega,t) \approx
\label{19}
\frac{1}{N}
\frac{\Delta\bbox{G}^{\omega}_{\bbox{k}}(\bbox{0},0) \otimes
 \Delta\bbox{G}^{\omega}_{\bbox{k}'}(\bbox{0},0)}{-i\Omega + 
\mu(\omega,t)+\bbox{K}\cdot \bbox{D}(\omega)\bbox{K}}
\end{equation}
with $N=-2\overline{n^3}\omega^2/(\pi c^3)$
and $\overline{n^3}$ being the angular and arithmetic average of the
two refractive indices.
The denominator represents a diffusion pole, which also contains an
``absorption'' coefficient $\mu(\omega,t)$. The diffusion tensor
follows from
\begin{equation}
\label{21}
\bbox{K} \cdot \bbox{D}(\omega) \bbox{K} =  \frac{c}{2\overline{n^3}} \,
[{\cal G}(\bbox{K})]^{\ast} \cdot {\cal B}^{-1} {\cal G}(\bbox{K}) 
\end{equation}
with
\[
 \protect[{\cal G}(\bbox{K})]_{\alpha i} =  \pi 
\int_{\hat{\bbox{k}}^{\alpha}} n_{\alpha}(\hat{\bbox{k}})
\left[ \frac{\partial \bbox{G}_0^{-1}}{\partial \bbox{k}}
  \bbox{K} \right]_{\alpha}\!\!\!\!(\hat{\bbox{k}}) \,
[\varphi_{i}^{\alpha}(\hat{\bbox{k}})]^{\ast}
\]
and
\begin{eqnarray*}
\protect[\,{\cal B}\,]_{\beta j}^{\alpha i} & = & 
\sum_{\gamma} \Bigl\{
 \pi\int_{\hat{\bbox{k}}^{\alpha}}\int_{\hat{\bbox{q}}^{\gamma}} 
\bigl[ [\varphi_{i}^{\alpha}(\hat{\bbox{k}})]^{\ast} 
\varphi_{j}^{\alpha}(\hat{\bbox{k}}) \delta^{\alpha}_{\beta}  \\
& &
-\, [\varphi_{i}^{\alpha}(\hat{\bbox{k}})]^{\ast} 
 \varphi_{j}^{\beta}(\hat{\bbox{q}}) \delta^{\gamma}_{\beta} \bigr] \, 
[\bbox{B}^{\omega}_{\bbox{k}^{\alpha}\bbox{q}^{\gamma}}(0)]_{\alpha \gamma}
\Bigr\}\enspace.
\end{eqnarray*}
In principal all $\varphi_{i}^{\alpha}(\hat{\bbox{k}})$ of odd parity 
contribute to $\bbox{D}(\omega)$.
For isotropic systems we choose spherical harmonics:
$\varphi_{i}^{\alpha}(\hat{\bbox{k}}) \rightarrow Y_{lm}(\vartheta,\varphi)$.
Only the components $[{\cal G}(\bbox{K})]_{\alpha l=1m}$ are nonzero and 
$[\,{\cal B}\,]_{\beta l'm'}^{\alpha lm} \propto \delta_{ll'}$. Therefore,
only spherical harmonics of $l=1$ contribute to $\bbox{D}(\omega)$ and we
get the familiar formular $D=\frac{1}{3}cl^{\ast} \propto [\langle \,
1-\text{cos}\vartheta \rangle]^{-1}$.
The absorption coefficient reads
\[
\mu(\omega,t) = \frac{c\pi^3}{2\overline{n^3}}
 \, \sum_{\alpha,\beta} 
\int_{\hat{\bbox{k}}^{\alpha}}\int_{\hat{\bbox{q}}^{\beta}} 
[ \bbox{B}^{\omega}_{\bbox{k}^{\alpha}\bbox{q}^{\beta}}(0)
-\bbox{B}^{\omega}_{\bbox{k}^{\alpha}\bbox{q}^{\beta}}(t)
 ]_{\alpha\beta} \enspace.
\]
It represents an angular average over all the dynamical modes of the system.
(For $t=0$, it is zero and then increases due to the decaying temporal
correlations in $\langle \, \delta\bbox{\varepsilon} \otimes 
\delta\bbox{\varepsilon} \, \rangle$.)
The numerator in Eq.\ (\ref{19}) indicates which initial and final
polarization states have a nonzero overlap with the diffusion pole.
The second factor $\Delta\bbox{G}^{\omega}_{\bbox{k}'}(\bbox{0},0)$ depends
only on the input wave. The first factor 
$\Delta\bbox{G}^{\omega}_{\bbox{k}}(\bbox{0},0)$ involves only the output
wave and determines the ratio of densities of photons in the two output 
polarization states 1 and 2 {\em independent\/} of the state of the input 
wave. An integration
over $k$ ($\int k^2 dk$) shows that this ratio is
$[n_1(\hat{\bbox{k}})/n_2(\hat{\bbox{k}})]^3$ for the wave direction 
$\hat{\bbox{k}}$. This effect should be measurable. 
Finally, the Green's function corresponding to $W_2(\bbox{R},t)$ follows from 
$\bbox{\Phi}^{\omega}_{\bbox{k}\bbox{k}'}(\bbox{K},\Omega=0,t)$ by 
integrating over $\bbox{k},\bbox{k}'$ and applying the appropriate trace 
operation.

The diffusion tensor $\bbox{D}(\omega)$ has the same uniaxial form as the
dielectric tensor in Eq.~(\ref{3}). We express the diffusion coefficients
$D_{\|}$ and $D_{\perp}$ in terms of a typical length $l_0^{\ast} = 9\pi\, 
\frac{c_{\perp}^2}{\omega^2}\,\frac{K_3}{k_{\text{B}}T}\, 
\frac{\varepsilon_{\perp}^2}{\Delta\varepsilon^2}$ 
($c_{\perp}=c/\sqrt{\varepsilon_{\perp}}$) times unitless numerical
factors $\widetilde{D}_{\|}$ and $\widetilde{D}_{\perp}$ via
\begin{equation}
\label{26}
D_{\|} = c_{\perp}l^{\ast}_0 \widetilde{D}_{\|} / 3 \enspace,
\enspace D_{\perp} = c_{\perp}l^{\ast}_0 \widetilde{D}_{\perp} / 3
\enspace,
\end{equation}
where $\widetilde{D}_{\|}$ and $\widetilde{D}_{\perp}$ depend on 
$K_1/K_3$, $K_2/K_3$ and $\Delta\varepsilon/\varepsilon_{\perp}$.
For the material 5CB, $K_3=5.3 \times 10^{-7} \,\text{dyne}$, 
$\varepsilon_{\perp}=2.381$ and $\Delta \varepsilon / \varepsilon_{\perp}
= 0.228$. With $T=300\,\text{K}$ and green light 
($\omega/c = 1.15\times 10^5 \,\text{cm}^{-1}$) we get 
$l_0^{\ast} = 2.3\,\text{mm}$, which is in agreement with experiments 
\cite{Vithana93,Yodh95}. 
As basis functions we choose spherical harmonics depending on a new
``polar angle'' $\vartheta'_{\alpha}$ 
(see \cite{Stark95c}). For the ordinary light ray
$\vartheta_2' = \vartheta$, for the extraordinary one 
$\vartheta'_{\alpha}$ is given by
$\text{cos}\vartheta'_1= n_1(\hat{\bbox{k}}) \text{cos}\vartheta /
\sqrt{\varepsilon_{\perp}}$. The basis functions
$\varphi^{\alpha}_{lm}(\hat{\bbox{k}}) = 
Y_{lm}(\vartheta'_{\alpha}(\vartheta), \varphi)$ are
orthogonal with respect to the weight 
$n_{\alpha}^3(\hat{\bbox{k}}) d\Omega_{\bbox{k}} \propto 
d\text{cos}\vartheta'_{\alpha} d\varphi$.
Then, only $[{\cal G}(\bbox{K})]_{\alpha l=1m}$ is nonzero \cite{Stark95c}.
We studied the contributions of different $l$ to $D_{\|}$ and $D_{\perp}$ and 
found that $l=3$ in addition to $l=1$ gives changes 
of less than $2\,\%$.
In Fig.\ \ref{fig:1}, we plot our results for $\widetilde{D}_{\|}$ and
$\widetilde{D}_{\perp}$ for 5CB with $K_1/K_3=0.79$, $K_2/K_3=0.43$ and
$\Delta \chi =0.95 \times 10^{-7}$. At $H=0$, $\widetilde{D}_{\|}=0.95$ and
$\widetilde{D}_{\perp}=0.65$ are finite even though, as noted earlier, the
scattering mean-free-path for the extraordinary light ray is infinite. The 
anisotropy in the diffusion 
constants decreases with both $\Delta \varepsilon$ and anisotropy in the Frank
elastic constants. In the limit $\Delta \varepsilon =0$ and $K_1=K_2=K_3$,
$D_{\|}/D_{\perp} = 1.06$ is not unity because of the inherent anisotropy
in the structure factor.
The diffusion approximation is valid only 
for times $t$ much smaller than characteristic relaxation times of the
director modes. In this case we get
\begin{equation}
\mu(\omega,t)\approx t \mu_0 \enspace, \enspace
\mu_0=\frac{2k_{\text{B}}T}{9\pi}\,\frac{\omega^4}{c_{\perp}^3}\,
\frac{\Delta\varepsilon^2}{\varepsilon_{\perp}^2} \,
\frac{\tilde{\mu}}{\gamma} \enspace,
\end{equation}
where $\gamma$ is the rotational viscosity and $\tilde{\mu}$ a numerical factor
depending on all other viscosities and 
$\Delta\varepsilon/\varepsilon_{\perp}$. Note that unlike scattering in 
colloids, $\mu_0$ depends only on viscosities and is independent of the
static structure factor ($\propto k_{\text{B}}T/Kq^2$). This is because 
the same fluctuations determine scattering and dynamics.
Finally, we point out that the appropriate
Laplace-Fourier transform of Eq.\ (\ref{19}) leads to a temporal 
autocorrelation function $W_2$ that can be expressed in a form reminiscent of 
the average over
light paths used in isotropic systems \cite{Maret87,Pine88}:
\begin{equation}
W_2 \propto \int_0^{\infty} d\tau \, P(\tau) \exp(-\mu_0 t \tau) \enspace,
\end{equation}
where $P(\tau)$ is the probability that an anisotropic random walker enters
the medium at a prescribed point and leaves it at another point after a time
$\tau$. (Note that this integral is over time $\tau$ rather than path length
because the light velocity is not a constant along an arbitrary path.)

This work was supported in part by the Deutsche 
For\-schungs\-gemeinschaft under Grant No. Sta\ 352/2-1 and by 
NSF under Grant No. DMR 91-20688. We wish to thank
Ming Kao, Kristen Jester and Arjun Yodh for helpful discussions.



\end{document}